# How Quantum is the Classical World?

Gary Bruno SCHMID[a] and Rudolf M. DUENKI[b]

[a] *Research Group F.X. Vollenweider, Psychiatric University Clinic, University of Zuerich, Lenggstrasse 31, CH-8029 Zuerich, Switzerland*
*E-mail: gbschmid@mac.com*
[b] *Dept. of Physics, University of Zuerich, Winterthurerstr. 190, CH-8057 Zürich, Switzerland*
*E-mail: rmd@physik.uzh.ch*

*Abstract*

It has been experimentally confirmed that quantum physical phenomena can violate the Information Bell Inequalities. A violation of the one or the other of these Information Bell Inequalites is equivalent to a violation of local realism meaning that either objectivity or locality, or both, do not hold for the phenomena under investigation. We propose (1) an experimental design for carrying out classical measurements in the absence of ontological complementarity; (2) a rational way to extract epistemologically complementary (pseudocomplementary) data from it; (3) a statistical approach which can reject stochastic and/or suspected violations of local realism in measurements of such data.

## 1   Introduction

The reader may remember asking themselves as a graduate student questions like:

> "Is a perfect, classical coin in an unstable balance between two states at once just before collapsing into a stable classical state of *Heads* or *Tails* actually in a quantum physical q-bit state, if only for a fraction of a second, before falling flat on a table top? And if so, could two such classical coins somehow become entangled?"

It's easy to ask this provocative question, but hard to design an experiment to falsify the corresponding Null Hypothesis $H_0$: *Local realism obtains globally*. (The concept of local realism is discussed below.) Nevertheless, quantum theory surrounding the quantum probability field does not exclude the possibility that the proposed null-hypothesis might not be correct. Unfortunately, however, the quantum probability field serves little more than as a wonderful black box, like the psychologist's unconscious and the believer's netherworld (German: Jenseits), within which rational explanations of all kinds of bizarre speculations can lie hidden unless an experiment can be designed and carried out to test them.

There are at least two difficulties involved in an experimental test of the above-mentioned null-hypothesis: (1) In contrast to quantum physics, there is, to our knowledge, no algebra capable of describing the above-mentioned or similar classical situations which can be used to test the Bell Inequality in a given case; (2) The utter absence of complementarity in classical measurements seems to make it impossible to design a classical experiment along the lines of a quantum experimental test of the Bell Inequality, even if such an algebra were to exist. Overcoming these difficulties is what this paper is all about.

The Bell expressions, be they standard or information-theoretic ones, involve correlations or probabilities which can be tested in the lab and allow experimental errors. As statistical statements, it is no wonder that, in a finite set of data, they can be violated by chance. If the violation happens rarely, the tested source can as well be local realistic. Indeed, and in spite of the provocative title, we do not wish to suggest that there has to be something "nonlocally realistic" going on in the classical world just because the violation of local realism can happen by chance using a classical model. Quite to the contrary, in this paper, we provide rigid criteria which would have to be fulfilled if a classical-like model could indeed – as shown here – now and again violate nonlocal realism.

## 2      Background

*Perspectives*

Bell's Inequality is a restriction placed by local realistic theories such as classical physics upon observable correlations between different systems in experiments. Bell's original theorem and stronger versions of this theorem „state that essentially all realistic local theories of natural phenomena may be tested in a single experimental arrangement against quantum mechanics, and that these two alternatives necessarily lead to observably different predictions (Clauser & Shimony 1978)." In this work, we study the probability to violate the Bell Inequality using two classical-like numerical models and a finite number of "experimental" runs. In the first model (stochastic case), the results of all local measurements are totally random and independent, also in different runs of the experiment. In the second model (anticorrelated case), the results for the settings chosen by the simulations are always anticorrelated, and all other predetermined results are random and independent. Note that these models are nothing more than numerical simulations of two possible, extreme, experimental outcomes. The results provide a statistical basis which can be used to estimate the significance of outcomes from real experiments such as the *quantum coin experiment* discussed below. It is only the experiments themselves – and not these numerical simulations - which, of course, must be expected to avoid certain loopholes which would otherwise allow violations of Bell's Inequality within the context of a classical experiment. Such loopholes include the locality, the fair-sampling, and the freedom-of-choice loopholes (Scheidl et al. 2009).

We also emphasize that although our experimental design might aid the explanation of possible observed violations of Bell's Inequality by local realistic theories, it's aim is to place a strong restriction on the acceptance of systematic violations of local realism in a particular set of repeated measurements. In order to do this, however, we have had to figure out a way to gather and analyse the results coming from a classical experiment which, of course and by its very nature, cannot involve ontologically complementary variables – see below.

### Information has its price

Quantum physics is the physics of the microcosmos, the world of tiniest things like molecules, atoms, electrons, protons, neutrons, photons etc.. The behaviour of things in the microcosmos obeys natural laws which seem to be in defiance of our trusted, everyday experience. This is especially true with regard to our empirically based belief in an objective, local reality. No one in their right mind, thinking in a normal, common sense way, would ever expect, for example, that two separated coins would always be able to display an opposite result: *heads* vs. *tails* or *tails* vs. *heads* with 100% reliability without their being some kind of information transfer - some kind of signal – taking place between them to correlate their outcomes nonlocally. But this is, of course, just the kind of behaviour which „quantum things" can – under particular experimental conditions - be shown to display.

There is still no way to logically get from the empirically established, nonlocal realism of the quantum world to the just-as-empirically-established local realism of the classical world in a self-consistent, complete, analytical way. This unsolved problem is often overlooked when trying to apply the ideas of quantum entanglement and teleportation as metaphors to understand certain classical observations such as unusual healing methods and psychological phenomena (Schmid 2005). For example: Although it is not unthinkable that the human mind in an especially prepared mental state of *psychological absorbtion, dream, flow, meditation or trance* etc. might somehow be able to biochemically isolate certain macromolecules from their thermodynamic environment within the brain, thus enabling them to partake of nonlocal, quantum phenomena, that is, to inhibit the decoherence of the Psi-function of the entire system (Zeilinger 1999, Zeilinger 2007), this non-impossibility is highly speculative and does not open the door to using quantum physics to simply explain such unphysical things as, say, mental telepathy, as many authors might like to believe.

Nature dictates that information has its price. The lowest non-negotiable one is defined by quantum physics: Carrying out a decision or expressing an intention, for example, measuring something, always disturbs something else, even if this disturbance might, as in most cases, go unnoticed. Let's take, for example, the famous double-slit experiment: Light enters a single, narrow opening from the left and illuminates a screen with two slits, each of which can be independently closed again. If both slits are left open, one observes interference bands on the screen attesting to the wave-like property of light. If one of these slits is closed, these bands disappear and one simply observes a single illuminated stripe, as if light were particle-like. Quantum interference only then occurs, if absolutely no information is available as to which way these light „particles" might might have taken. The point is not whether or not an observer might actually be in possession of this information but, rather, whether or not they could in any possible way, even if only in principle, deduce which way these particles might have taken. In order for interference to be observed, it must be impossible for anyone – no matter where they might be and no matter what kind of sophisticated technology they might have at their disposal – to figure out which of the two possible ways these light particles have taken. In other words, for the wave-like interference to occur, one must be able to completely and totally isolate the system from its (thermodynamic) environment so that no kind of surreptitious information transfer, like, say, the emission of an electromagnetic or some other kind of signal, might, in principle at least, be able to give a clue as to which path the light particles are taking.

### *Braunstein & Caves Experiment*

Consider the experiment suggested by Braunstein & Caves (B & C) consisting of two counter-propagating spin-1/2 particles, *A* and *B* (Braunstein & Caves 1988). These particles are emitted by the decay of a zero-angular-momentum particle and, accordingly, have spins $S_A$ and $S_B$ such that $S_A + S_B = 0$. Each particle is sent through a Stern-Gerlach apparatus which measures a component of the particle's spin along one of two possible directions labelled by the unit vectors **a** and **a'** for particle *A*, and **b** and **b'** for particle *B*. Accordingly, for particle *A* there are two observables A = $S_A$ · **a** and A' = $S_A$ · **a'** . Similarly, for particle *B* there are two observables B = $S_B$ · **b** and B' = $S_B$ · **b'** . For a spin ½ system, the possible values of A and A' labelled by a and a', respectively, and denoted by the quantum numbers: m = +1/2 and m = -1/2 are dichotomous, that is, a = +½ or -½ and a' = +½ or -½. (States with positive or negative spin values are called *spin-up* or *spin-down* states, respectively.) Accordingly, there are 4 observables: A and A' associated with *A*, and B and B' associated with *B*.

*Entanglement* of two systems means that, at any given instant of time after their separation and isolation, the overall, combined system is simultaneously in two states at once, both of which correspond to two observably different outcomes for the individual systems. For example, consider the overall, combined system of two spinning coins: $Coin_1$ and $Coin_2$ being tossed about in a shaker. After letting them fly separate ways out of the shaker, the combined system could be thought of as being simultaneously in the following four states at once: ($Coin_1$-spin-up and $Coin_2$-spin-down) AND ($Coin_1$-spin-down and $Coin_2$-spin-up) AND ($Coin_1$-spin-up and $Coin_2$-spin-up) and ($Coin_1$-spin-down and $Coin_2$-spin-down). All four states in ()'s correspond to four observably different outcomes for the individual coins. In the jargon of quantum physics, one says that the state vector of the overall, combined system is a *vector of the second type* (Selleri & Tarozzi 1981).

In the experiment of B & C, we have the following condition of entanglement: $S_A + S_B = 0$. This relation entangles the spins A and B of particles *A* and *B* into an ambiguously superposed state of the form: (A-spin-up and B-spin-down) AND (A-spin-down and B-spin-up). The same pertains for A' and B'. Such an ambiguous, superposed state is called a *q-bit*. Furthermore, in a quantum physical system, the two observables X and X' associated with each system *X* do not commute under certain conditions, for example in this case, when the angle between the unit vectors **x** and **x'** is 90 degrees. Under this condition, X and X' cannot be observed simultaneously, that is, the values of x and x' can not be determined simultaneously. This *condition of complementarity (non-commutativity)* between X and X' in each system leaves corresponding "holes", that is, empty cells, in the data table – see Table 2. It is just this combination of entanglement for a certain preparation of states and complementarity for a certain range of angles between **x** and **x'** which leads to violations of the peculiar statistics underlying Bell's Inequality. In the given case of a system with spin zero disintegrating spontaneously into two spin-1/2 particles, that is, to the zero-angular-momentum state of $S_A$ and $S_B$, this leads to a ca. 41% degree of violation of Bell's Inequality for **x** perpendicular to **x'**, with an angle of $135^0$ between **a** and **b'**, and $45^0$ between both **b'** and **a'** as well as between **a'** and **b** (Selleri & Tarozzi 1981, p. 15).

The factors involved in the Stern-Gerlach experiment are summarized in Table 1.

| Table 1 Particle | Physical quantity being measured (Stern-Gerlach apparatus) | Observables | Measurement values (quantum numbers) | Physical laws |
|---|---|---|---|---|
| *A* | Spin angular momentum: $S_A$ | A = $S_A \cdot$ **a**  <br> A' = $S_A \cdot$ **a'** | a = +½ or -½  <br> a' = +½ or -½ | **Entanglement:** $S_A + S_B = 0$ <br> This condition entangles the spins A and B or A' and B' of particles *A* and *B*. <br> **Complementarity:** <br> Let X denote either A or B. Then X & X' do not commute and x and x' are not simultaneously measureable for **x** ⊥ **x'**. <br> This condition leaves "holes" in the data table – see Table 2. |
| *B* | Spin angular momentum: $S_B$ | B = $S_B \cdot$ **b** <br> B' = $S_B \cdot$ **b'** | b = +½ or -½ <br> b' = +½ or -½ | |

**Table 1.** Factors involved in the Stern-Gerlach Experiment.

Assume a Stern-Gerlach experiment is carried out at an arbitrary angle $0° < \Theta < 180°$ between the unit vectors **a** and **b** specifying the orientations of the apparatus. For simplicity and as suggested by B & C, we require **a'** and **b'** to lie between **a** and **b** within the plane defined by **a** and **b**, that is, all four vectors are coplanar. The angles between **a**, **b'**, **a'** and **b** are fixed and, for simplicity, are taken to be $\Theta/3$. Each outcome or "run" of the experiment results in a total of 2 observable events leading to a pair of values: one value for either a or a' and one value for either b or b'. Because of the noncommutativity between A and A' as well as between B and B', values for a and a' cannot be determined simultaneously and the same is true for b and b'. Accordingly, each outcome consists of one or another of the 16 = (2x2)x(2x2) possibilities shown in Table 2.

| Table 2 Outcome | a | a' | b | b' |
|---|---|---|---|---|
| 1 | 1 | --- | 1 | --- |
| 2 | 1 | --- | 0 | --- |
| 3 | 1 | --- | --- | 1 |
| 4 | 1 | --- | --- | 0 |
| 5 | 0 | --- | 1 | --- |
| 6 | 0 | --- | 0 | --- |
| 7 | 0 | --- | --- | 1 |
| 8 | 0 | --- | --- | 0 |
| 9 | --- | 1 | 1 | --- |
| 10 | --- | 1 | 0 | --- |
| 11 | --- | 1 | --- | 1 |
| 12 | --- | 1 | --- | 0 |
| 13 | --- | 0 | 1 | --- |
| 14 | --- | 0 | 0 | --- |
| 15 | --- | 0 | --- | 1 |
| 16 | --- | 0 | --- | 0 |

**Table 2.** All 16 possible outcomes of a spin-½ Stern-Gerlach experiment. Here, the values 0 and 1 correspond to spin values of – ½ and + ½, respectively.

Ideally, one would record the results from, say, 100 or more such outcomes so as to obtain a distribution amongst the various cells of the frequency cross table between the values for a, a' and b, b' as shown in Table 3.

| Table 3 | | a=-1/2 | a=+1/2 | a'=-1/2 | a'=+1/2 |
|---|---|---|---|---|---|
| | | *N(a=-1/2)* | *N(a=+1/2)* | *N(a'=-1/2)* | *N(a'=+1/2)* |
| **b=-1/2** | *N(b=-1/2)* | N(a=-1/2,b=-1/2 \| Θ) | N(a=+1/2,b=-1/2 \| Θ) | N(a'=-1/2,b=-1/2 \| Θ/3) | N(a'=+1/2,b=-1/2 \| Θ/3) |
| **b=+1/2** | *N(b=+1/2)* | N(a=-1/2,b=+1/2 \| Θ) | N(a=+1/2,b=+1/2 \| Θ) | N(a'=-1/2,b=+1/2 \| Θ/3) | N(a'=+1/2,b=+1/2 \| Θ/3) |
| **b'=-1/2** | *N(b'=-1/2)* | N(a=-1/2,b'=-1/2 \| Θ/3) | N(a=+1/2,b'=-1/2 \| Θ/3) | N(a'=-1/2,b'=-1/2 \| Θ/3) | N(a'=+1/2,b'=-1/2 \| Θ/3) |
| **b'=+1/2** | *N(b'=+1/2)* | N(a=-1/2,b'=+1/2 \| Θ/3) | N(a=+1/2,b'=+1/2 \| Θ/3) | N(a'=-1/2,b'=+1/2 \| Θ/3) | N(a'=+1/2,b'=+1/2 \| Θ/3) |

**Table 3.** Schematic, empirical frequency distribution amongst the possible outcomes of a spin-1/2 Stern-Gerlach experiment. The frequencies depend only on the angle Θ between a and b, and the angle Θ/3 between a, b', a', and b. The total number of observable events per experiment $N_{exp}$ is given by the sum of all 16 cross terms in this table. *Accordingly, all statistics derived from this table are based on $N_{exp}$/(2 observable events per outcome) = $N_{outcomes/experiment}$.*

Quantum physics provides an algebra for calculating the theoretical values in the cells of a second cross table – see Table 4 - of corresponding conditional probabilities p(a|b | Θ), p(a|b' | Θ), p(a'|b | Θ), p(a'|b' | Θ) for a given angle Θ (Braunstein & Caves 1988, p. 663).[1]

| **Table 4** | **a=-1/2** | **a=+1/2** | **a'=-1/2** | **a'=+1/2** |
|---|---|---|---|---|
| **b=-1/2** | p(a=-1/2:b=-1/2 \| Θ) | p(a=+1/2:b=-1/2 \| Θ) | p(a'=-1/2:b=-1/2 \| Θ/3) | p(a'=+1/2:b=-1/2 \| Θ/3) |
| **b=+1/2** | p(a=-1/2:b=+1/2 \| Θ) | p(a=+1/2:b=+1/2 \| Θ) | p(a'=-1/2:b=+1/2 \| Θ/3) | p(a'=+1/2:b=+1/2 \| Θ/3) |
| **b'=-1/2** | p(a=-1/2:b'=-1/2 \| Θ/3) | p(a=+1/2:b'=-1/2 \| Θ/3) | p(a'=-1/2:b'=-1/2 \| Θ/3) | p(a'=+1/2:b'=-1/2 \| Θ/3) |
| **b'=+1/2** | p(a=-1/2:b'=+1/2 \| Θ/3) | p(a=+1/2:b'=+1/2 \| Θ/3) | p(a'=-1/2:b'=+1/2 \| Θ/3) | p(a'=+1/2:b'=+1/2 \| Θ/3) |

**Table 4.** Schematic quantum theoretical, conditional probability cross table of predicted results.

There are several important things to point out here:

1. The quantum physical algebra determines the distribution of frequencies ultimately resulting in the conditional probabilities given by the p(a|b | Θ)'s.

   If we didn't have this algebra, the frequency cross table (Table 3) would have to be used to define Table 4 with cells containing the conditional probabilities calculated as p(a|b)=N(a,b)/N(b) etc.. The empirical joint probabilities p(a,b)=N(a,b)/($N_{outcomes/experiment}$) which can also be deduced from Table 3 are not needed. Here, $N_{outcomes/experiment}$ is equivalent to the number of experimental outcomes or "runs" defining a given experiment, in this case, 16. In general, $N_{outcomes/experiment}$ = $N_{exp}$/(number of observable events per outcome). Classically, a cross table of joint probabilities would also be constructed from Table 3, and these joint probabilities would then be used to derive the conditional probabilities from Bayes' Theorem as shown below. Note, however, that in the classical case there are no empty cells so that the number of observable events per outcome is twice as large as in the quantum case: *Accordingly, all statistics derived from the classical Table 3 must be based on $N_{outcomes/experiment}$ = $N_{exp}$/(4 observable events per outcome)*.

2. The physics of the experimental design determines that the angle of the argument of the p(a|b | Θ)'s in the 4 a x b cells is Θ and in the remaining 12 a x b', a' x b, and a' x b' cells is Θ/3.

3. The quantum physical p(a|b | Θ)'s will necessarily *not* obey Bayes' Theorem in the range of angles Θ within which there is a positive[2] information deficit – see Fig. 1 of B & C - and will not *necessarily* obey Bayes' Theorem outside this range (Braunstein & Caves 1988, p. 663).

   To verify this, one would need to calculate the joint probabilities p(a,b) etc. from the frequency table of the original data as mentioned above under Point 1. In the paper of B & C, no experiment was carried out, so that no original data table exists from which such a frequency cross table can be constructed.

4. Physics determines the complementarity between the cells a and a' or b and b'. This results in missing data in the nonobservable cells.

---

[1] Notice that the notation in Braunstein & Caves is somewhat different. The conditional probabilities in their Eqn. 10 use commas to separate the arguments, thus making it easy to mistake them for joint probabilities.
[2] We prefer to use the negative of the information difference defining the information deficit in Eqn. 12 of Braunstein & Caves.

## 3  Purpose & Design

***Pseudocomplementary Data: Ontological versus Epistemological Complementarity***

Now what if we don't know the algebra underlying a phenomenon, but we do have a series of measurements at hand and want to test the original data table for violations of local realism? (This situation was briefly mentioned under Point 1 above.) If the phenomenon involves complementarity, there will be missing data in the cells complementary to the measurable ones. Let us call this the case of *strong* or *ontological complementarity*. *Strong complementarity* obtains in the quantum world. In this case, we would simply proceed as outlined above to calculate the information deficit from only those frequency cells providing data.

But what if complementarity only obtains in a *weak* or *epistemological* sense (Atmanspacher et al. 2002) ? Could there be a clever way to select the one or the other variable, a or a', from system **A**, and b or b' from system **B** so as to result in a violation of local realism? In other words, what if there would be some way to know in each outcome which of the two measurements, a or a', from system **A**, and b or b' from system **B**, is physically relevant, even though the complementary quantity in each system is also observable (but, presumably, not physically relevant)? This knowledge would take over the role of the complementarity between certain observables which leads to the noncommutativity in the algebra we always have in the quantum physical case. Then the particular set of selected measurements of the one quantity a or a' in system **A** and b or b' in system **B** from each outcome might reliably result in a violation of local realism whereas the overall set of all "weakly" complementary quantities from both systems would simply yield random results which may or may not stochastically violate local realism.

Let us assume that the data selected in some clever way as mentioned above is the "real", physically relevant data. (Exactly how this data is collected, whether objectively according to some algorithm or device or, subjectively according to intuition, need not concern us for the purposes of this paper.) We call such a data set: *pseudocomplementary data* and the remaining data – which would otherwise be missing because it would not be observable in a real quantum experiment – *hidden data*. Note that the hidden data are extracted from the basic population in a fashion similar to the extraction of the pseudocomplementary data, but are assumed to involve little to no "entanglement". *Accordingly, the entropies are additionally conditioned on the pseudocomplementary data as explained in Table 6 below in order to use epistemological complementarity to reconstruct or mimic ontological complementarity. This statistical mimicry is the gist of our analysis.*

The pseudocomplementary data can be understood to lead to the p(a|b | Θ/3)'s in Table 4.

***How Quantum is the Classical World?***

A classical phenomenon taken together with an algorithm to reliably carry out a clever choice of psedocomplementary data could be thought to be quantum physical at the extent to which the data resulting from this algorithm violates local realism. The B & C Inequality involves conditional entropies between measurement results and, as such, is fully operational and can be established in any theory to test violations of local realism. The question is now: *How can we discover whether or not local realism is violated in the physically relevant results, that is, in the pseudocomplementary data of a classical phenomenon? In other words, how quantum is the classical world?*

***Quantum Coin Study Design***

In an attempt to answer the above questions, we carried out the following two extreme *quantum coin computer simulations* under selection of only one of the two variables, a and a', from system **A,** and similarly for system **B**. (We will return to this *quantum coin simulation* again further below within the context of a corresponding gedankenexperiment.) Accordingly, the counterpart to the quantum physical results schematically shown in Tables 2 and 3 above is obtained by constructing pseudocomplementary data for a new Table 2 in the following ways:

1. *Stochastic Case:* For the given outcome, randomly generate, four times, a 0 ("Tails") or a 1 ("Heads"), and write each of these values, one after the other, into the corresponding column of data: a, a', b, b'. This procedure will gradually lead to a full data matrix of random results with no empty cells.

   For each outcome, as mentioned above, select one column from each of the groups (a,a') and (b,b'), for example, a and b' (instead of a and b, or a' and b, or a' and b') according to some selection rule, e.g., by means of a random number generator or intentionally by a clever, educated guess. This defines the subset of *pseudocomplementary data* – in this example, cells a and b' - with the remaining, so-called "hidden data" located in the remaining, unselected cells – in this example, cells a' and b.

   For a given outcome, all results are random: the selection of the values: 0 or 1 as well as the choice of complementary observables.

2. *Anticorrelated Case:* For each outcome: randomly (or cleverly – see above) define one of the columns a or a' and, similarly, only one of the columns b or b' to contain the pseudocomplementary data. (The other cells, by default, contain what we have come to call *hidden data* values in this outcome.)

   Now randomly generate each of the pseudocomplementary values a or a' as a 0 ("Tails") or a 1 ("Heads"). Then require the value of the corresponding pseudocomplementary cell b or b', respectively, to be anticorrelated with the pseudocomplementary cell a or a', and the value of the remaining hidden cell to be random. Accordingly, all selected (pseudocomplementary) cells have perfect anticorrelation between them, and all nonselected (hidden) cells contain random results. Again we have a full data matrix (=no empty cells) whereby only half of the data is assumed to be physically relevant, namely, the pseudocomplementary data.

A computer simulation or "experiment", for short, was defined to consist of 4, 8, 12, or 16 pseudocomplementary numerical outcomes. A set of 10'000 experiments was carried out in each case for each of the two above-mentioned extremes: Stochastic Case and Anticorrelated Case. The information deficit, H_Deficit_Pseudo, was calculated for each computer experiment as suggested in the paper of B & C in the following way:

$$H\_Deficit\_Pseudo = H(A|B \mid hd) - \{H(A|B' \mid pd) + H(B'|A' \mid pd) + H(A'|B \mid pd)\} \text{ bits} \qquad (1)$$

where the abbreviations, hd and pd, stand for *hidden data* and *pseudocomplementary data*, respectively. (We will have more to say about the definition of H_Deficit below in the section on Bayes' Theorem.)

Note that B & C use two different statistical distributions in the Information Bell Inequality to analyze an information deficit in their simulated Stern-Gerlach experiments. To make this most clear, consider the case with $\Theta=90^0$: The presumably entangled distribution for angle $\Theta/3 = 30^0$ between the corresponding pairs of unit vectors (**a**, **b'**), (**b'**, **a'**), and (**a'**, **b**) is used to calculate all three terms $\{H(A|B')+H(B'|A')+H(A'|B)\}$, and the presumably non-entangled distribution for angle $\Theta=90^0$ between the corresponding pair of unit vectors (**a**, **b**) is used to calculate the single term H(A|B). (See their Eqn. (12).) *Accordingly we similarly use the presumably entangled distribution pd to calculate all three terms $\{H(A|B')+H(B'|A')+H(A'|B)\}$ and the presumably non-entangled distribution hd to calculate the single term H(A|B).*

If H_Deficit_Pseudo>0, then four objective quantities whose statistics are extracted from joint probabilities derived from the pseudocomplementary data carry less information than any two objective quantities whose statistics are extracted from the joint probabilities derived from the hidden data. If this is the case, we prefer not to use the expression *entanglement* but, rather, suggest using a terminology borrowed from depth psychology by speaking here of a *synchronicity* between events.

Mathematically, the largest possible value for H_Deficit_Pseudo is H(A | hd). This is the case when, in the pseudocomplementary data, all the information about A is already contained in B' AND all the information about B' is already contained in A' AND all the information about A' is already contained in B AND when, in the hidden data, no information about A is contained in B. Accordingly, the expression

$$\text{Index\_Deficit} = \text{Max}(\text{H\_Deficit\_Pseudo})/\text{H}(A \mid hd) \qquad (2)$$

is a measure of the extent to which nonlocal realism is violated in a set of experiments. Here, the value of H(A | hd) for the data matrix with maximum H_Deficit_Pseudo is used. Another measure which also proves useful is

$$\text{Index\_Norm} =$$

$$\text{Max}(\text{H\_Deficit\_Pseudo}) / \{\text{Max}(\text{H\_Deficit\_Pseudo}) - \text{Min}(\text{H\_Deficit\_Pseudo})\} \qquad (3)$$

A MS-EXCEL programm for calculating the above quantities within automated sets of 10'000 experiments is available from the authors upon request.

## 4     Results

We are interested in the two cases mentioned above:

1. Random selection of pseudocomplementary cells and random results in cells
2. Random selection of pseudocomplementary cells and perfect anticorrelation between results within the selected cells

In both cases, it is helpful to know from a given set of experiments:

- the probability that the value of H_Deficit_Pseudo is greater than zero: p(H_Deficit_Pseudo>0)
- the frequency of values of H_Deficit_Pseudo greater than zero: No(H_Deficit_Pseudo>0)
- the average value of H_Deficit_Pseudo: Avg(H_Deficit_Pseudo)
- the maximum value of H_Deficit_Pseudo: Max(H_Deficit_Pseudo)
- Index_Deficit
- Index_Norm

On the one hand, it is reasonable to assume that especially in the first extreme *quantum coin experiment* described here, the value of H_Deficit_Pseudo will always be less than zero since, in this stochastic case, there is no essential difference in the statistical nature of the hd and pd data sets. Nevertheless as shown in Table 7, this condition is, indeed, sometimes, if only rarely, violated by chance. On the other hand, in the second, anticorrelated case, one might expect local realism to be greatly influenced by the significant mathematical differences in the statistical correlations within the hd and pd data sets. However, also under this condition is local realism only seldomly violated.

The results for repeating a set of 10'000 computer experiments with experiments comprising only four, eight, or twelve outcomes per experiment are presented in Table 5.

| Table 5<br>n= No. outcomes/exp.<br>N = No. valid experiments | p(H_Deficit_Pseudo>0)<br>$n_0$(H_Deficit_Pseudo>0)[3] | Avg(H_Deficit_Pseudo>0) | Max(H_Deficit_Pseudo) | Index_Deficit<br>Index_Norm |
|---|---|---|---|---|
| n = 4<br>N = $10^4$ | 0.348 (random)<br>$n_0$=933 (random)<br><br>0.483 (anticorrelation)<br>$n_0$=1327 (anticorrelation) | 0.19 bits (random)<br>0.21 bits (anticorrelation) | 0.50 bits (random)<br>0.50 bits (anticorr.) | 1.000 (random)<br>0.327 (random)<br><br>1.000 (anticorr.)<br>0.400 (anticorr.) |
| n = 8<br>N = $10^4$ | 0.056 (random)<br>$n_0$=443 (random)<br><br>0.157 (anticorrelation)<br>$n_0$=1198 (anticorrelation) | 0.11 bits (random)<br>0.12 bits (anticorrelation) | 0.50 bits (random)<br>0.50 bits (anticorr.) | 1.000 (random)<br>0.317 (random)<br><br>1.000 (anticorr.)<br>0.413 (anticorr.) |
| n = 12<br>N = $10^4$ | 0.012 (random)<br>$n_0$=108 (random)<br><br>0.052 (anticorrelation)<br>$n_0$=500 (anticorrelation) | 0.07 bits (random)<br>0.09 bits (anticorrelation) | 0.30 bits (random)<br>0.41 bits (anticorr.) | 0.698 (random)<br>0.225 (random)<br><br>0.823 (anticorr.)<br>0.362 (random) |
| n = 16<br>N = $10^4$ | 0.003 (random)<br>$n_0$=25 (random)<br><br>0.024 (anticorrelation)<br>$n_0$=237 (anticorrelation) | 0.07 bits (random)<br>0.08 bits (anticorrelation) | 0.21 bits (random)<br>0.34 bits (anticorr.) | 0.429 (random)<br>0.181 (random)<br><br>0.711 (anticorr.)<br>0.326 (anticorr.) |

**Table 5.** Results of 10'000 computer experiments with both random- and anti-correlation between pseudocomplementary cells for various numbers n of outcomes per experiment.

The major results from Table 5 are:

1. The smaller the number n of outcomes per experiment, the greater the probability {p(H_Deficit>0) & $n_0$(H_Deficit>0)} that this set of outcomes will lead to a positive information deficit.

2. The smaller the number n of outcomes per experiment, the greater the *average value of H_Deficit_Pseudo* for a given degree of correlation between the results of the pseudocomplementary cells. Nevertheless, even in the case of perfect anticorrelation for n=4, the average value of H_Deficit_Pseudo is only 0.21 bits.

3. The smaller the number n of outcomes per experiment, the greater the *maximum value of H_Deficit_Pseudo* for a given degree of correlation between the results of the pseudocomplementary cells. Nevertheless, even in the case of perfect anticorrelation for n=4, the maximum value of H_Deficit_Pseudo is only 0.50 bits.

4. For a given number n of outcomes per experiment, the *average value of H_Deficit_Pseudo* is larger for perfect as opposed to random correlation between results in the pseudocomplementary cells. Similarly, for a given number n of outcomes per experiment, the *maximum value of H_Deficit_Pseudo* for perfect correlation between results in the pseudocomplementary cells is greater or equal to that for random correlation.

---

[3] The p(H_Deficit_Pseudo>0) = 1.0-PERCENTRANK(H_Deficit_Pseudo values;0) values deviate from the n-values for H_Deficit_Pseudo>0 because of redundancies (0's) in the H_Deficit_Pseudo data and the way EXCEL treats redundancies when evaluating the function PERCENTRANK: PERCENTRANK(0) = (No. values < 0)/{(No. values <0)+(No. values =0)+(No. values >0)}. Such redundancies start occurring for n<12 outcomes/experiment.

5. The smaller the number of outcomes n per experiment, the less the extent (Index_Deficit & Index_Norm) to which nonlocal realism (H_Deficit_Pseudo>0) is violated has to do with the degree of correlation between the pseudocomplementary data.

Our pseudocomplementary classical data mimic a quantum physical spin ½ system. In the quantum physical spin ½ system, the spins are perfectly anticorrelated under entanglement. In Figure 1 of B & C we see that although the value of Max(H_Deficit_Pseudo) for the quantum physical systems under consideration increases with spin number from ca. 0.25 bits for spin ½ to ca. 0.45 bits for spin 25, the physical window (angle between entangled vectors) within which entanglement occurs shrinks to 0 (Braunstein & Caves 1988, p. 664). Other authors have shown that *although local realism can indeed be violated by noise-resistant systems of arbitrarily high dimensionality d, the quantum state in question vanishes at the extent (probability) to which it is affected by noise* (Collins et al. 2002, eqn. 20). Casually summing up "on the back of an envelope": *the larger the spin, the closer the system is to being classical, but the greater is the probability that the system is affected by noise and the smaller is the physical window within which entanglement can be expected to occur.* Furthermore, in the quantum physical case under the experimental condition of entanglement, any and every set of outcomes, no matter how large, always leads to a positive information deficit: p(H_Deficit_Pseudo>0) = 1.00. This is the case, for example, for all angles Θ in the range between 0 and the crossing point of the information difference curve with the null-information-difference axis in Figure 1 of B & C.

### *Distributions of H_Deficit_Pseudo with both random and anticorrelation between pseudocomplementary cells for n = 4 and 16*

The distributions of H_Deficit_Pseudo from Eqn. (1) are given for n=4 and n=16 in Figure 1 and 2, respectively, and are labelled *H_D_Hidden* in reference to the first term H(A|B | hd) in this equation. (Distributions for n=8 and n=12 are available from the authors upon request.) The average covariances for all experiments as well as for only those experiments leading to positive values of H_Deficit_Pseudo are shown in the respective figure captions. (Covariances instead of correlations are given for randomly generated data because EXCEL gives singular results for data with correlations of +1.0.)

**Figure 1.** Distribution of H_Deficit_Pseudo with both (a) random and (b) anticorrelation between pseudocomplementary cells for n = 4.

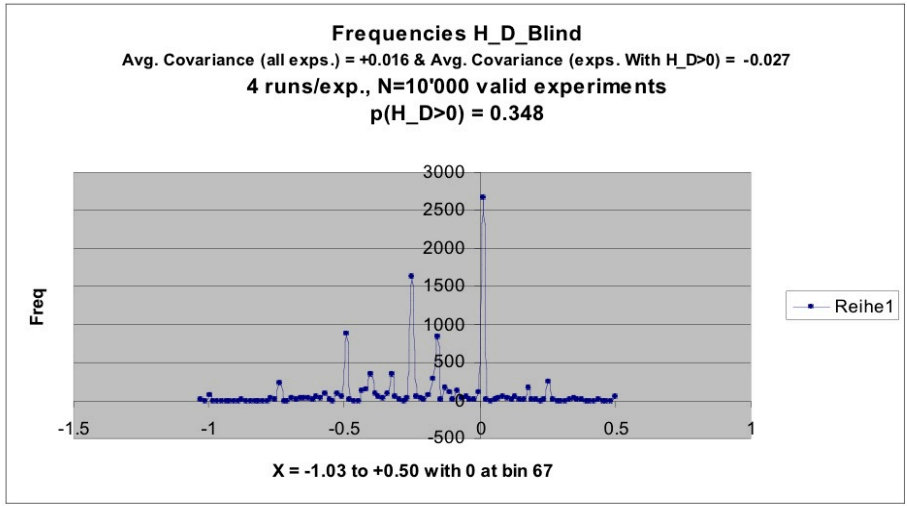

(a) Random Correlation

(b) Anticorrelation

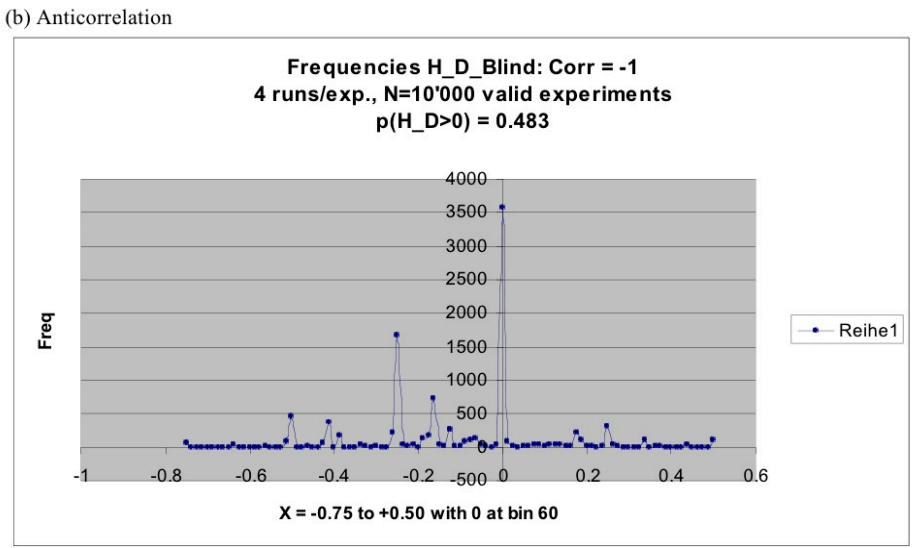

**Figure 2.** Distribution of H_Deficit_Pseudo with both (a) random and (b) anticorrelation between pseudocomplementary cells for n = 16.

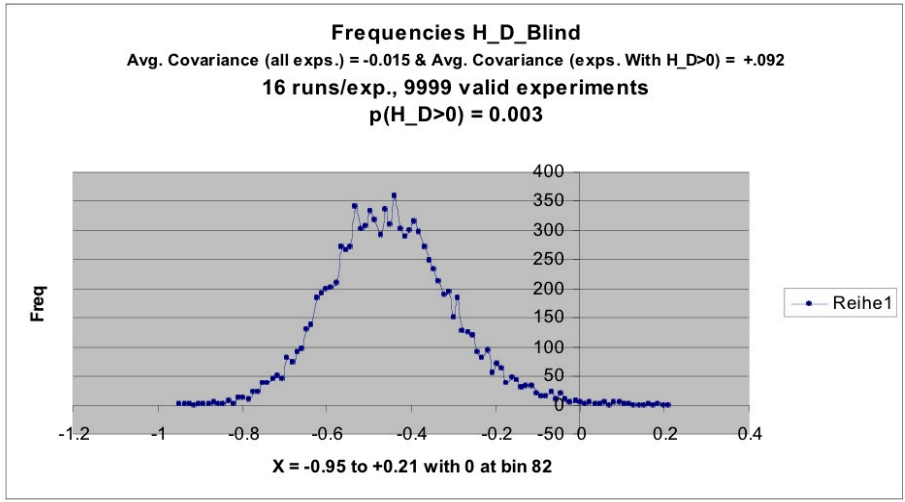

(a) Random Correlation

(b) Anticorrelation

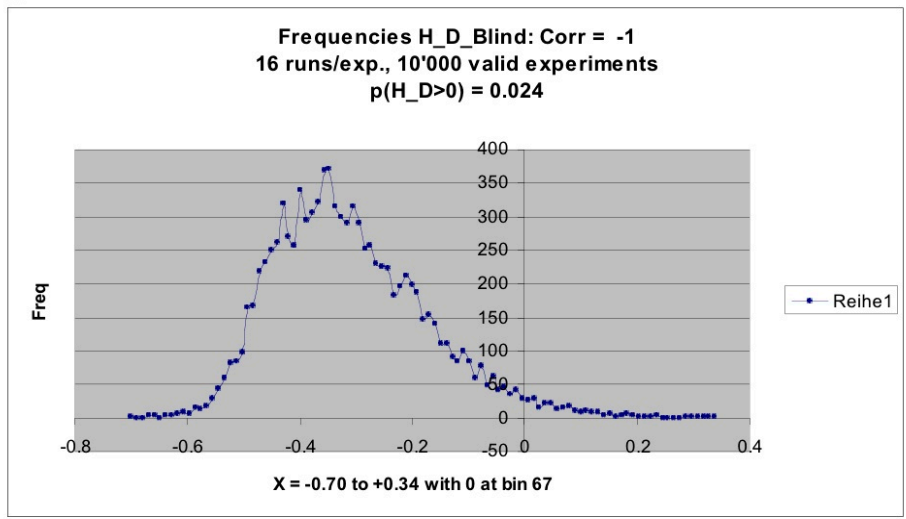

***Statistical Significance of Violations of Local Realism***

*We have demonstrated that local realism can be violated stochastically even in classical experiments when carried out in the above-mentioned way to accomodate pseudocomplementary data. The results of our computer simulations also offer a way to test for a possible violation of local realism onhand an appropriate statistical measure for the goodness of classical violations of local realism.*

From the standpoint of inference statistics, it is advisable to à priori deny the violation of local realism and to state this denial as the Null Hypothesis $H_0$: *Local realism obtains globally*. This means that any violations would be regarded to be non-deterministic (stochastic), i.e., to occur only randomly. The acceptance of a systematic violation in repeated measurements is then the Alternative Hypothesis $H_1$: *Local realism does not obtain in a particular set of repeated measurements*.

Let us assume the null-hypothesis. The probability that a spin s system displays a positive H_Deficit may then be directly read from the sample of outcomes of the corresponding simulation: This probability, $P_{random}[H\_Deficit > \delta \mid H_0]$ or $p_0$ for short, is just the ratio of the number of outcomes displaying H_Deficit $> \delta$ ($\delta >= 0$), #(H_Deficit $> \delta$), divided by the size, #(simulations), of the whole sample:

$$P_{random}[H\_Deficit > \delta \mid H_0] = \#(H\_Deficit > \delta)/\#(simulations) := p_0 \qquad (4a)$$

The probability $P_{random}[H\_Deficit > \delta \mid H_0]$ in (4a) actually designates the probability that H_Deficit $> \delta$ occurs spontaneously, i.e. randomly.

Now let a Monte Carlo experiment be conducted N times and let $k_e$ be the number of times we find H_Deficit $> \delta$. Accordingly, $k_e$ is a statistically observed value. The (random) probability *under the null-hypothesis* for at least $k_e$ positive outcomes is just[4]

$$P_{ke}[H\_Deficit > \delta \mid H_0] = 1 - \Sigma_k \, N!/((N-k)! \, k!) \, p_0^k \, (1 - p_0)^{N-k} \, , \, 0 \leq k < k_e \qquad (4b)$$

Note that the probability $P_{ke}[H\_Deficit > \delta \mid H_0]$ in (4b) represents the probability that H_Deficit $> \delta$ occurs $k_e$ times stochastically and this is generally different from (4a). If this probability (4b) is larger than a certain prespecified level α (i.e. α = 0.05, 0.01, 0.001), the random result given by (4a) is not significant: one accepts the null-hypothesis and rejects the alternative. On the other hand, if this probability is smaller than a certain prespecified level α (i.e. α = 0.05, 0.01, 0.001), one rejects the null-hypothesis and accepts the alternative. (Note that it is not clear how one could practically determine $P_{ke}[H\_Deficit > \delta \mid H_0]$ without carrying out a Monte Carlo simulation here.)

We note that inference statistical reasoning works by hypothesis exclusion: the acceptance of $H_1$ follows from the rejection of the null-hypothesis and not by ´proving` the validity of $H_1$. In the absence of a quantitavely specified hypothesis, this is the type of test which can be (repeatedly) carried out. (Most psychological, medical and biostatistical experiments are probably of this nature.) Alternatively one could extract the outcome from an actual experiment as suggested below in the section "Pseudocomplementary Quantum Coin Gedankenexperiment" and determine $P_{experiment}[H\_Deficit > \delta]$ from (4a) – because one can't know à priori whether or not such an experiment involves an underlying physical process or simply results from random events - and then reject the null-hypothesis if this result is larger than α. But then this would not be an inference statistical criterium and the reasoning could no longer be carried out with the help of eqns. (4a) and (4b).

---

[4] The sum in this formula assumes a discrete distribution of values for k. Equation (4b) can also be understood to express the probability of finding H_Deficit $> \delta$ *more* than $k_e$ out of n times.

**Number of Required Experiments $N_{req}$**

If – as is the case in quantum physical experiments – a quantitatively specified Alternative Hypothesis $H_1$ is formulated (see below), one can determine the number of experiments $N_{req}$ which need to be conducted to be able to accept the Null Hypothesis, $H_0$ with a significance level of $(1-\alpha)$, or $H_1$ with a significance level of $\gamma$ ($\gamma <\sim 1.0$). However, in the case that the Null Hypothesis $H_0$ is rejected because $\alpha$ is, say, 6%, it would be mistaken to assume that the Alternative Hypothesis $H_1$ can be accepted at the level of 94%. In other words, we still need to know to what extent the rejection of $H_0$ means that the Alternative Hypothesis $H_1$ can be accepted at some level, say, $\gamma \geq 80\%$.

On the one hand, we require under the null-hypothesis, as in the case above, that the probability of a certain (or a more extreme) outcome in violation of the null-hypothesis should not exceed $\alpha$. (Usually, we take $\alpha \leq 0.01$.) On the other hand, we demand that, under the alternative hypothesis, the probability for such or a more extreme outcome is at least $\gamma$. (Usually, we take $\gamma \geq 0.80$.) Both these requirements are expected to be fulfilled within the theoretical limit defined by the value $k_0$. In other words for an arbitrary number N, one has to find some $k_0$ such that for k outcomes with H_Deficit > 0:

$$P_N^{Ho} ( k > k_0 ) < \alpha\% \quad (5a)$$

AND

$$P_N^{H_1} ( k > k_0 ) \geq \gamma\% \quad (5b)$$

, that is, both $H_0$ would be rejected AND $H_1$ would be accepted. In the binomial case, the cumulated probability $P_N$ may be written for each hypothesis $H_x$ (x = 0 or 1) as

$$P_N^{Hx} (k > k_0) = 1 - \Sigma_k \, N!/((N-k)! \, k!) \, (p_0^{Hx})^k (1 - p_0^{Hx})^{N-k} \, , \, 0 \leq k < k_0 \quad (6)$$

Holding N fixed, $k_0$ is the minimal number of positive outcomes satisfying the above relations (5). *The least N for which such a $k_0$ exists is then the minimum number of experiments required $N_{req}$.*

If one now conducts $N_{req}$ experiments and finds $k_e$ outcomes with H_Deficit > $\delta$ and $k_e > k_0$, the Alternative Hypothesis $H_1$ is accepted. Algorithms to solve this problem are normally implemented as functions in the more powerfull statistical software packages (e.g. the function 'Power and Sample Size' in S-Plus). The following gedankenexperiment based on the *quantum coin computer experiment* discussed earlier helps explain this. (See Table 7 below.)

**Pseudocomplementary Quantum Coin Gedankenexperiment**

Consider a physicist having a kind of classical Stern-Gerlach apparatus which, as far as he knows, may or may not be secretely manipulated in a way to be described below. He can adjust this apparatus to measure the information deficit in the experiment discussed by B & C for their Fig.1. His experiment consists of two systems **A** and **B** which, after being sent through the detector, evidence one of two values which, for simplicity, we will call *Heads* or *Tails* along one of two possible directions labelled by the unit vectors **a** and **a'** for system *A*, and **b** and **b'** for system *B*. However, he is faced with a tricky problem in this gedankenexperiment, namely, he can't be sure whether or not the values he gets have been secretely manipulated: The results he gets might have been generated either by a random number generator – for example, by the toss of a coin - or by the actual physics of a measurement process involving spin-½ particles and their respective complementary observables. This is because the tricky apparatus fills in – with random values – cells which would otherwise remain empty in a fully quantum experiment. (In other words, the cells with "---" in Table 2 would contain randomly selected 0's and 1's.) *Accordingly, the experimentor has to decide from the statistics of the experimental outcomes alone whether or not he is observing the behaviour of a real, quantum physical system or that of a simple random number generator.*

All our experimenter can do is use his gut feeling and guess which measurements are physical and which can be neglected as artifactual. For reasons which should already be clear from our discussion of the *quantum coin computer experiment* above, the "complementarity" in this gedankenexperiment is not so much between measurements as it is between the mental choices made by the experimentor as to which two of an overall set of four measurements he intuitively decides to define to be the "physical" ones.

Our experimentor doesn't know whether or not he is dealing here with a fully quantum phenomenon and, therefore, doesn't have any algebra to help him rationally select the angles between the unit vectors: **a**, **a'**, **b** and **b'**. Nevertheless, since he knows that, if the phenomenon is quantum at all, it involves spin-½ particles, he decides for simplicity to also select these angles as discussed above: all four vectors coplanar, $\text{angle}_{ab}=\theta$, $\text{angle}_{ab'}=\theta/3$, $\text{angle}_{b'a'}=\theta/3$, $\text{angle}_{a'b}=\theta/3$. Then, he picks the angle $\theta$ between the unit vectors **a** and **b** at random between 0 and 100 degrees.

Notice that this gedankenexperiment as well as the *quantum coin experiment* discussed below can be carried out to satisfy the assumptions of *realism* – definite values exist for all variables defining the state of objects prior to and independent of observation -, *locality* – space-like separated events cannot causally influence each other - and *freedom-of-choice* – the choice of measurement settings is free or random - required by Bell to derive his inequality.

The gist of this gedankenexperiment is outlined schematically in Table 6:

| Table 6<br>**Analysis Steps** | **Classical Physics** | | **Quantum Physics** |
|---|---|---|---|
| | **Purely Classical** | **Pseudocomplementary** | **Purely Quantum** |
| **Collecting raw data**<br><br>N outcomes per experiment<br><br>Two systems:<br>Each system has one pair of complementary observables<br>Four observables:<br>Two pairs of complementary observables<br><br>(See Table 1) | Full data matrix<br><br>(Like Table 2 but with experimental values and no empty cells)<br><br>All cells regarded equal.<br><br>No selection carried out | Full data matrix<br><br>(Like Table 2 but with experimental values and no empty cells)<br><br><br><br>*Selection of pseudocomplementary cells* | Incomplete data matrix (empty cells) due to complementary values<br><br>(Like Table 2 but with experimental values and some empty cells)<br><br><br>Selection unnecessary because of ontologically complementary data |
| **Frequency cross table** | Frequency cross table (Table 3) from data in all cells | *Frequency cross table (Table 3) from data in selected cells only.*<br>*Hidden data is used for the frequencies in the 4 upper-left quadrants.*<br><br>*Pseudocomplementary data is used for the frequencies in the remaining 12 quadrants.* | Frequency cross table (Table 3) from data in non-empty cells only.<br>Complementary data collected at Θ is used for the frequencies in the 4 upper-left quadrants.<br>Complementary data collected at Θ/3 is used for the frequencies in the remaining 12 quadrants. |
| **Intermediate calculation of joint probabilities only for the full classical case** | Table of joint probabilities p(a, b) from Table 3 based on 4 events per outcome | | |
| **Calculation of conditional probabilities** | Table 4 of conditional probabilities calculated using the p(a, b)-joint-probability table with Bayes' Theorem | Table 4 of conditional probabilities calculated directly from the frequency cross table (Table 3) | Table 4 of conditional probabilities calculated directly from the frequency cross table (Table 3) |
| **Calculation of H_Deficit from the conditional probabilities** | H_Deficit $\geq 0^5$ never, using Eqn. (7) in B & C | *H_Deficit $\geq$ 0 sometimes, using Eqn. (1)* | H_Deficit $\geq$ 0 sometimes, using (the negative of) Eqn. (12) in B & C |

**Table 6.** Schematic illustrating the differences in the steps involved in calculating the information deficit for the purely classical case, the pseudocomplementary classical case, and the purely quantum case.

As his null-hypothesis, our experimenter assumes he is dealing with a classical world, namely, with the random flip of a coin in every case. In other words, he assumes that local realism obtains. However, he does know from Figure 1 of B & C that, for spin-1/2 particles behaving quantum physically, the probability for the Alternative Hypothesis $H_1$ is $p_0^{H1}$ =0.85 (assuming a uniform distribution of angles between 0° and 100° of the Stern-Gerlach apparatus). *His situation is, however, not identical to the one described above in the section entitled „Braunstein & Caves Experiment" since he can, indeed, measure all four quantitites a, a', b and b' in each outcome. Nevertheless, he also obtains the frequency distribution shown above schematically in Table 3.*

---

[5] Recall Footnote 2 above regarding the sign of H_Deficit.

Assume our experimenter intends to carry out a series of experiments whereby a single experiment comprises n=12 outcomes. From our pseudocomplementary results shown in Table 5 above, he reads the probability $p_0^{Ho} = 0.012$ (if he assumes the results are randomly correlated) or $p_0^{Ho} = 0.052$ (if he assumes the results are anticorrelated). From Figure 1 of B & C he reads $p_0^{H1} = 0.85$ (assuming a uniform distribution of angles between 0° and 100° of his Stern-Gerlach apparatus). Inserting Equation (6) into Eqns. (5) enables him to determine the minimum number of experiments $N_{req}$ together with the minimum number of violations $k_0+1$ of the null-hypothesis necessary to accept the alternative hypothesis for a particular goodness of result as defined by the pair [α, γ]. (Recall Eqns. (5).) The result is displayed in Table 7.

**Table 7**

| α [%] | γ [%] | $N_{req}$ | $k_0$ |
|---|---|---|---|
| 5 | 80 | 3 | 0 |
| 1 | 80 | 3 | 1 |
| 0.5 | 80 | 3 | 1 |
| 0.1 | 80 | 3 | 1 |
| 5 | 90 | 3 | 0 |
| 1 | 90 | 4 | 1 |
| 0.5 | 90 | 4 | 1 |
| 0.1 | 90 | 4 | 1 |
| 5 | 95 | 4 | 0 |
| 1 | 95 | 4 | 1 |
| 0.5 | 95 | 4 | 1 |
| 0.1 | 95 | 4 | 1 |
| 5 | 99 | 4 | 0 |
| 1 | 99 | 5 | 1 |
| 0.5 | 99 | 5 | 1 |
| 0.1 | 99 | 6 | 2 |

**Table 7.** The minimum required number of trials $N_{req}$ for the gedankenexperiment with different levels of α and γ in the random case ($p_0^{Ho} = 0.012$). If within $N_{req}$ experiments, the outcome H_deficit > 0 is not found for more than $k_0$ times, the Null Hypothesis $H_0$ is not rejected. If this result is found more than $k_0$ times, the Alternative Hypothesis $H_1$ is accepted and is valid within the specified significance. Only a few outcomes are required to determine the correct hypothesis with high accuracy.

The low values for $N_{req}$ in Table 7 are due to the fact that, in the case under study, the one probability is very small: $p_0^{Ho}=0.012$ and the other one is very large $p_0^{H1}=0.85$. If one were to try to distinguish, say, a random correlation: $p_0^{Ho} = 0.012$ from an anticorrelation: $p_0^{H1}=0.052$ for [α = 0.01, γ = 0.95], the minimum number of required experiments $n_{req}$ would already be 312 and $k_0 = 10$. For $k_e > k_0$ experiments with H_Deficit > δ, we would then assume that the pseudocomplementary data are significantly anticorrelated.

To use this approach, one must have estimates of $p_0^{Ho}$ and $p_0^{H1}$ to generate - from Eqn. (6) - a corresponding table equivalent to Table 7. A concrete example using Table 7 for the purposes of this paper is discussed in the Conclusions section below.

<span style="color:blue">In the case of a psychological experiment there is no algebra similar to that available to B & C, i.e. there is no quantifiable $H_1$. In other words, the psychologist is doomed to simply carry out a standard test of significance as indicated above in the section "Statistical Significance of Violations of Local Realism" with $p_0^{H1}$ less than or equal to, say, 0.012. This is because, in this case, $p_0^{H1}$ is not available.</span>

## 5    Discussion

*This paper offers a method by which one can determine whether or not a set of seemingly random data – which displays a well-defined, built-in complementarity structure – is "classically random".* This kind of analysis might be especially relevant to the field of quantum cryptography (QCY). QCY is interested in deciding whether or not a message has been eavesdropped by determining whether or not the statistical structure of complementarity inherent to the data set has been disturbed. This can be formulated more concretely as follows: *For a given data set, do the statistics of entangled events follow a classical statistics – then one has been eavesdropped – or a quantum statistics – then the message was not eavesdropped.* The point is that any and every act of eavesdropping destroys the quantum physical correlations between entangled events.

Here we do something quite similar although more general. *Instead of offering statistical tests for the comparison: "classical versus quantum statistics", we offer a comparison "classical versus nonclassical statistics" WITHOUT the explicit assumption of a quantum physical background. This is certainly of interest because a deviation per se would already be a surprising result independent of knowing the details of the underlying mechanism.*

### *Objectivity, Locality and Local Realism*

*Objectivity is a kind of reality assumption.*[6] It means that, at any given instant in time, all physical quantities considered to be state variables have definite values, independent of observation. Specifying values to all state variables (classical physics) or quantum numbers (quantum physics) at any given instant in time uniquely defines the momentary physical state of a system.

> Thus, both classically and quantum physically, objectivity means that an object is always in a definite physical state independent of observation.

Vice versa, at any given instant in time, forcing a system into any one of its possible physical states means uniquely defining values to all its state variables (classical physics) or quantum numbers (quantum physics). Accordingly, in each outcome of an experiment, all measurable physical quantities have definite values independent of observation. In classical physics, this is always the case for all state variables. In quantum physics, complementary (=mathematically noncommuting) physical quantities cannot be measured simultaneously within a given system. However, - and that's the claim of objectivity - it can be assumed even in the case of a complementary quantity that, in principle at least, the value of the "hidden" variable which is not determined and, hence, remains unknown, is nevertheless well-defined.

This "in principle at least" is the crux of the matter surrounding the statistical implications of objective realism. Statistically speaking, objectivity means that the statistics of outcomes that measure a set of physical quantities are given by corresponding joint and conditional probabilities. *Joint probabilities* define the likelihood of finding simultaneous (=joint) values for two or more measurable quantities, called *observables*, some belonging to the one system, some to the other. Under the condition that one already has knowledge about one or more of these values for the one system, there will be a certain probability, called a *conditional probability*, that one will discover certain values for some of the variables associated with observables of the other system. The more the two systems in question share information between themselves (=so-called *mutual information*), the larger this conditional probability will be on the average (and vice versa).

*Locality is a kind of "nondisturbance assumption".* This means that if two systems are isolated, a measurement on one does not disturb the results of any measurements on the other (Braunstein & Caves 1988, p. 663). The idea of locality goes hand-in-hand with the idea of *local causes*, that is, separate systems can only influence one another via currents of substance-like quantities (mass-energy, momentum, angular momentum, entropy, charge etc.) flowing between them continuously through interstitially neighboring volumes of space-time (Schmid 1984).

---

[6] What we call *objectivity* is also referred to as *realism* by other authors. See, e.g., (Clauser & Shimony 1978).

Statistically speaking, locality means that the statistics of outcomes that measure a pair of physical quantities, one associated with each of two isolated systems, are entirely defined by the corresponding joint probabilities, in this case, by pair probabilities. In the case of isolated systems, the joint probabilities are given by simple products of corresponding single probabilities.

*Local realism assumes the validity of both objectivity and locality, that is, that physical systems have objective, local properties. Local realism* always fulfills two assumptions: (1) Locality, i.e., no „substance" can disappear at one place and reappear at another without having flowed, continuously, through the interstitial regions of space separating them; (2) Objectivity, i.e., this substance exists as a physical quantity unceasingly at each and every point in time between its moving from the one place to the other, even if its existence has not been verified along the way.

Statistically speaking, local realism establishes the existence and relevance of the relationship between the joint and the conditional probabilities according to Bayes' Theorem. With the help of Bayes' Theorem, the world picture of local realism dictates how any two systems must carry information between themselves and, in this way, constrains the statistics of measurements on two presumably separated systems.

From the point of view of developmental psychology, an infant reaches conscious objectivity when he or she can realize: "My mother exists even when I don't see or hear her behind the door!", and conscious locality when the infant can realize: "My mother can only then be aware of my changing needs when I communicate these to her; she can't just 'know' them!" It seems that objectivity is psychologically more basic, i.e. is acquired earlier in development, than locality. At any rate, beyond a certain age, local realism is a matter of course for every (normal) child: Whereas for the very young child, their mother might seem to have been "destroyed" at the one side of a dividing wall only to be "created" again at the other side, the slightly older child will understand their mother after disappearing to be existing the whole time while out of sight and walking continuously from the one region to the next behind the wall until she finally reaches the other side where she reappears again.

### *Bayes' Theorem & the Information Bell's Inequality*

To explain Bayes' Theorem, consider the case of two isolated systems **A** and **B**, each possessing a corresponding measurable quantity (observable) A and B with possible values labelled by a and b, respectively. Then we have

$$p(a,b) = p(a|b)p(b) = p(b|a)p(a) \tag{7}$$

where p(a,b) is the joint probability of finding both a and b simultaneously and p(a|b) is the probability of finding value a for A if one has already obtained value b for B (and similarly for p(b|a)). This relation is the well-known *Bayes' Theorem*.

Classical statistics based upon Bayes' Theorem allows one to deduce the *Information Bell Inequality* (Braunstein & Caves 1988, eqn. (7), p. 663)

$$H(A|B) \leq \{H(A|B')+H(B'|A')+H(A'|B)\} \tag{8}$$

for a given set of data. This and similar constraints have been known in the literature as Information Bell Inequalities (Bell 1964), (Clauser & Shimony 1978), (Michler et al. 1996). *If local realism holds for the dynamics associated with the two systems at hand, then these systems must carry information consistent with these inequalities. A violation of the one or the other of these Information Bell*

*Inequalites is equivalent to a violation of local realism meaning that either objectivity or locality, or both, do not hold in the phenomena under investigation.*

*It has been experimentally confirmed that quantum physical phenomena can violate the Information Bell inequalities.* Accordingly, an information deficit can be defined from Eqn. (8):

$$H\_Deficit = H(A|B) - \{H(A|B')+H(B'|A')+H(A'|B)\} \qquad (9)$$

As already mentioned, a violation of the Information Bell Inequaltiy, that is, a positive value for H_Deficit – recall Footnote 2 - means that four objective quantities carry less information than any two of them. The logical, physical conclusion from this is that local realism has been violated, that is, either all four quantities do not have definite values independent of observation (violation of objectivity), or the pair probabilities are not given by simple products of corresponding single probabilities (violation of locality), or both (violation of Bayes' Theorem).

Due to the peculiar algebra of quantum mechanics, the usual assumption under empirical violations of local realism is that, primarily, the condition of locality does not hold and physicists say that the systems are entangled. (The idea of objectivity seems psychologically harder to give up on – see above.) It is tempting to adhere to this tradition and vernacular, assuming that empirical violations of the Information Bell Inequalites - should they ever be discovered to be statistically significant in a classical experiment - imply a violation of locality. Accordingly, significant empirical violations of the Information Bell Inequalites would, in principle at least, allow one to speak of entanglement as possibly being responsible for the evidence. However, since a violation of local realism within a classical experimental design as suggested here does not rigorously require that quantum physical entanglement be involved in the phenomena at hand, we prefer, in such cases, to use the depth-psychological term *synchronicity*, as already mentioned above.

### Quantum Coin Experiment

A real, that is, classical coin is, so to say, a quantum object with infinite spin, meaning that $p_0^{H1}$ approaches zero in Eqn. (6). (The reader can also infer this asymptotic behaviour from the progression of ever narrower curves in the *information difference vs. degrees* plot of Figure 1 in the B & C article.) As a consequence, equations (5) and (6) would require an $N_{req}$ approaching infinity to satisfy any of the goodness criteria [α, γ] shown in Table 7 to prove entanglement between a pair of classical objects. On the other hand, one could argue that the spin number attributed to the object might just as well inherently depend upon the nature of the observation (measurement process) as it does upon the nature of the object itself. In this case, somehow intuitively "knowing" which of two real measurements on an infinite spin, classical object is the physically relevant one could be understood to reduce its observable behaviour to, say, that of a spin-1/2 quantum object (if the outcome Table 3 for the pseduocomplementary data were to be identical to the outcome Table 3 from real measurements on an actual spin-1/2 quantum object. In this case, the use of $p_0^{H1} = 0.85$ as above would be justifiable.)

Now assume that a psychologist would use a real coin in a *quantum coin study design* in which 12 outcomes are considered to comprise a single experiment. (Recall the gedankenexperiment above.) From Table 5, we find $p_0^{Ho} = 0.012$. If he wants to arrive at results which, according to Table 7, violate Eqn. (7) with a goodness of [α = 0.001, γ = 0.99], then his design must allow for $N_{req}=6$ experiments to complete the data set for a single investigation. In this case, $k_0+1=3$ experiments must result in a positive H_Deficit_Pseudo in order to take his claim of entanglement seriously, that is, $k_e$ must be at least 3. As just mentioned above, this would mean that his method of mental observation, that is, his intuitive way of cleverly selecting the pseudodata, has somehow been able to mentally reduce the spin-number behaviour of the presumably entangled coins down to the lesser complexity of low-spin (quantum) objects.

In the words of Selleri and Tarozzi (Selleri & Tarozzi 1981), the triumphal successes of quantum theory in explaining the world of atoms and molecules and, to a lesser extent, nucleons and elementary particles, "constitute by themselves a heavy argument against a realistic conception of Nature: a physicist who has full confidence in quantum mechanics cannot maintain that atomic and subatomic systems exist objectively in space and time and that they obey causal laws." And these physicists had an algebra to back up these seemingly parapsychological or paraphysical ideas! Perhaps one could say that a theoretical physicist trying to apply quantum physical thinking to classical situations without a corresponding algebra is in even greater danger of entering the world of "quantum psychosis". With the ideas based upon quantum physics as developed in this paper, we offer a method to avoid this danger of undisciplined thought while mathematically investigating the possibility of entanglement/synchronicity phenomena in an otherwise obviously classical system.

## 6     Conclusion & Outlook

*Pseudocomplementary data collected in the way suggested here from a real experiment can be tested for the statistical significance of an initially hypothesized synchronicity between the (dichotomous) values of two physical quantities measured in two isolated, classical systems. Such a synchronicity would indicate a possible violation of local realism.* Computer experiments show that local realism can be violated stochastically in the case of classical measurements using an experimental design mimicring ontological complementarity. The more outcomes a given experiment encompasses, the less probable it is that the data from this experiment will violate local realism by chance. The degree to which local realism is violated by chance is relatively independent of the correlation of results between the outcomes of a given experiment. *This paper offers a way to test suspected classical violations of local realism for statistical significance.*

## 7     Acknowledgments

No research support was involved in this work.

# 8     Table & Figure Captions

**Table 1.** Factors involved in the Stern-Gerlach Experiment.

**Table 2.** All 16 possible outcomes of a spin-½ Stern-Gerlach experiment. Here, the values 0 and 1 correspond to spin values of – ½ and + ½, respectively.

**Table 3.** Schematic, empirical frequency distribution amongst the possible outcomes of a spin-1/2 Stern-Gerlach experiment. The frequencies depend only on the angle $\Theta$ between **a** and **b**, and the angle $\Theta/3$ between **a**, **b'**, **a'**, and **b**. The total number of observable events per experiment $N_{exp}$ is given by the sum of all 16 cross terms in this table. *Accordingly, all statistics derived from this table are based on $N_{exp}/(2$ observable events per outcome$) = N_{outcomes/experiment}$.*

**Table 4.** Schematic quantum theoretical, conditional probability cross table of predicted results.

**Table 5.** Results of 10'000 computer experiments with both random- and anti-correlation between pseudocomplementary cells for various numbers n of outcomes per experiment.

**Table 6.** Schematic illustrating the differences in the steps involved in calculating the information deficit for the purely classical case, the pseudocomplementary classical case, and the purely quantum case.

**Table 7.** The minimum required number of trials $N_{req}$ for the gedankenexperiment with different levels of α and γ in the random case ($p_0^{Ho} = 0.012$). If within $N_{req}$ experiments, the outcome H_deficit > 0 is not found for more than $k_0$ times, the Null Hypothesis $H_0$ is not rejected. If this result is found more than $k_0$ times, the Alternative Hypothesis $H_1$ is accepted and is valid within the specified significance. Only a few outcomes are required to determine the correct hypothesis with high accuracy.

**Figure 1.** Distribution of H_Deficit_Pseudo with both (a) random and (b) anticorrelation between pseudocomplementary cells for n = 4.

**Figure 2.** Distribution of H_Deficit_Pseudo with both (a) random and (b) anticorrelation between pseudocomplementary cells for n = 16.